\documentclass[iop,apj]{emulateapj}

\newcommand{\inv}{{\sc Invers10}}
\newcommand{\teff}{$T_{\rm eff}$}
\newcommand{\logg}{$\log g$}
\newcommand{\vsini}{$v_{\rm e}\sin i$}
\newcommand{\kms}{km\,s$^{-1}$}
\newcommand{\hd}{HD\,37776}
\newcommand{\he}{\ion{He}{1}~5876~\AA}
\newcommand{\bz}{$\langle B_{\rm z} \rangle$}

\newcommand{\figps}[1]{\resizebox{\hsize}{!}{\rotatebox{0}{\includegraphics{#1}}}}
\newcommand{\fifps}[2]{\centering\resizebox{#1}{!}{\includegraphics{#2}}}

\newcommand{\mybf}[1]{{#1}}

\shorttitle{Magnetic field of the helium-strong star HD\,37776}

\shortauthors{Kochukhov et al.}

\begin{document}

\title{The extraordinary complex magnetic field of the helium-strong star HD\,37776}


\author{Oleg Kochukhov\altaffilmark{1}, Andreas Lundin\altaffilmark{1}, Iosif Romanyuk\altaffilmark{2}, and Dmitry Kudryavtsev\altaffilmark{2}}
\altaffiltext{1}{Department of Physics and Astronomy, Uppsala University, Box 516, Uppsala 75120, Sweden; oleg.kochukhov@fysast.uu.se}
\altaffiltext{2}{Special Astrophysical Observatory of RAS, Nizhnij Arkhyz 369167, Russia; roman@sao.ru, dkudr@sao.ru}

\begin{abstract}
The early-type chemically peculiar stars often show strong magnetic fields on their surfaces. These magnetic topologies are organized on large scales and are believed to be close to an oblique dipole for most of the stars. In a striking exception to this general trend, the helium-strong star HD\,37776 shows an extraordinary double-wave rotational modulation of the longitudinal magnetic field measurements, indicating a topologically complex and, possibly, record strong magnetic field. Here we present a new investigation of the magnetic field structure of HD\,37776, using both simple geometrical interpretation of the longitudinal field curve and detailed modeling of the time-resolved circular polarization line profiles
with the help of magnetic Doppler imaging technique.
We derive a model of the magnetic field structure of HD\,37776, which reconciles for the first time all magnetic observations available for this star. We find that the local surface field strength does not exceed $\approx$\,30 kG, while the overall field topology of HD\,37776 is dominated by a non-axisymmetric component and represents by far the most complex magnetic field configuration found among early-type stars.
\end{abstract}


\section{Introduction}
\label{intro}

A sizable fraction of the main sequence A and B stars are observed to be strongly magnetic, with typical surface field strengths between a few hundred G and more than 30~kG \citep{donati:2009}. Unlike the active late-type stars, in which tangled magnetic topologies are generated by a contemporary dynamo involving differential rotation and turbulence, the early-type stars most likely acquire their magnetic fields in the process of star formation and then retain them during the main sequence evolution \citep[e.g.,][]{mestel:2003}. To maintain stability, these fossil fields must have a mixed poloidal-toroidal geometry in the stellar interior \citep{braithwaite:2006}, but at the surface the fields are expected to be essentially poloidal, with the dipolar contribution dominating the field geometry.

Observations of the mean line of sight magnetic component, the so-called longitudinal magnetic field, \bz, generally reveal a smooth, single-wave variation with the stellar rotation \citep{mathys:1991,bychkov:2005,auriere:2007}, consistent with a simple, dipolar-like structure of the magnetic field topologies in the early-type stars. Recent comprehensive investigations of the linear and circular polarization spectra suggested the presence of small-scale magnetic field structures in some chemically peculiar A-type stars \citep{kochukhov:2004d,kochukhov:2010}. At the same time, these studies still found a dipolar-like large-scale field organization, just as one would expect from simple \bz\ curves of these stars.

Only a few early-type stars show longitudinal magnetic field variation deviating significantly from a single-wave sinusoid. These stars, \hd\ \citep{thompson:1985}, \mybf{HD\,32633 \citep{leone:2000}}, HD\,133880 \citep{landstreet:1990}, HD\,137509 \citep{kochukhov:2006c}, and $\tau$~Sco \citep{donati:2006b}, attracted considerable attention because even a relatively small distortion of the \bz\ phase curve indicates a dominant higher-order multipolar field component. This challenges theoretical models to find a satisfactory explanation for the existence and apparent stability of these rare global stellar magnetic field topologies. In this context, investigations of unusual non-dipolar fields represent an important step towards achieving a more complete theoretical understanding of the hot-star magnetism.

Among the stars with conspicuously non-dipolar longitudinal magnetic field variation, \hd\ (HIP\,26742, V901~Ori) has the most bizzare \bz\ curve. This young He-strong B2IV star belongs to the Orion OB1b association \citep{landstreet:2007} and shows a prominent rotational variation of the brightness in different photometric bands \citep{adelman:1997b}, spectral line intensities \citep{pedersen:1977,pedersen:1979}, and flux distribution \citep{adelman:1985} due to the combined effect of a strong magnetic field and an inhomogeneous distribution of chemical elements over the stellar surface. Uniquely for an early-type star, its longitudinal magnetic field curve shows four well-defined extrema during each rotation cycle.

A consistent rotational period of $P_{\rm rot}\approx1.5387$~d follows from the variation of the longitudinal magnetic field \citep{thompson:1985}, the strength of the neutral He lines \citep{pedersen:1979}, and multi-color photometry \citep{adelman:1997b}. \citet{mikulasek:2008} presented a combined analysis of different time-series observations obtained for \hd\ over 31 years. They reported a remarkably rapid rotational spin down by 17.7~s during this period.

The first quantitative model aiming to explain the puzzling longitudinal magnetic field curve of \hd\ was proposed by \citet{bohlender:1994}. The \bz\ measurements were reproduced with an axisymmetric combination of the dipolar, quadrupolar and octupolar magnetic components with the strengths 3.4, $-59$, and 44~kG respectively. The resulting mean surface magnetic field strength exceeds the 34~kG field intensity measured for Babcock's star \citep{babcock:1960}, making \hd\ the most magnetized object among all non-degenerate stars. However, this conclusion is sensitive to the adopted parameterization of the field geometry and assumed inclination of the stellar rotational axis. The very large predicted field strength cannot be confirmed by direct detection of the Zeeman splitting of spectral lines due to \mybf{the} rapid rotation of \hd.

\citet{khokhlova:2000} derived magnetic field geometry and surface maps of chemical elements for \hd\ with the help of the Doppler-Zeeman imaging technique. The magnetic field structure was approximated by a superposition of non-aligned dipole and quadrupole components. The study employed a simplified treatment of the line formation in \mybf{a} magnetic field and was based on the circular polarization spectra with a sparse phase coverage. Despite these shortcomings, the theoretical $uvby$ light curves calculated for the surface abundance distributions of \citet{khokhlova:2000} fit the observations very well \citep{krticka:2007}. On the other hand, no definite magnetic field model could be found in \citeauthor{khokhlova:2000} study, with lines of different elements giving $B_{\rm d}=10$--80~kG, $B_{\rm q}=40$--160~kG, and different orientations of the magnetic axes. None of these models succeeded in reproducing the observed longitudinal magnetic field curve.

The inconclusive results of the previous magnetic field studies of \hd\ and the lack of a definite magnetic field geometry model compatible with different types of magnetic observations warrants a new investigation of this unique star. In this paper, we present a new analysis of the magnetic field topology of \hd\ using all magnetic field observations available for this star. In Section~\ref{obs}, we describe the longitudinal magnetic field measurements and time-resolved spectropolarimetric observations employed in our analysis. Sections~\ref{multipolar} and \ref{mdi} provide a description of the multipolar and Doppler imaging modeling of the magnetic observations. In Section~\ref{3dfld} we study the magnetospheric structure of \hd\ using potential field extrapolation. The paper concludes with the discussion of our results in Section~\ref{disc}.

\section{Observational data}
\label{obs}

\subsection{Longitudinal magnetic field measurements}

A strong longitudinal magnetic field in \hd\ was found by \citet{borra:1979}. Their seven \bz\ measurements showed the presence of a kG-strength variable magnetic field, which exhibits a peculiar double-wave modulation with rotation phase. Longitudinal magnetic field measurements of \hd\ were continued by \citet{thompson:1985}. All these observations were carried out using a photopolarimetric Zeeman analyzer, tuned \mybf{to} the wings of \mybf{the} H$\beta$ line. In total, 44 photopolarimetric \bz\ measurements were collected in 1977--1978 and in 1982--1984, spanning the range from $-2.2$ to $+2.5$~kG with a typical uncertainty of 300--400~G.

In 1993, five additional spectropolarimetric longitudinal field measurements were obtained for \hd\ by \citet{donati:1997}. These authors derived the line of sight magnetic field component from a mean Stokes $V$ profile of 17 metal absorption lines. The errors of these measurements are 170--360~G. Despite being of a rather different nature compared to the H$\beta$ photopolarimetric longitudinal field diagnostic, the \bz\ data points published by \citet{donati:1997} agree well with the previously established longitudinal field phase curve. Taking this into account, we combined the photo- and spectropolarimetric \bz\ determinations in the analysis presented below.

We recalculated the rotation phases of the longitudinal magnetic field measurements using the cubic ephemeris of \citet{mikulasek:2008}, which takes into account the rapid rotational braking of \hd. The updated longitudinal magnetic field curve is presented in Figure~\ref{fig:bz}. In this plot the phase 0.0 corresponds to the maximum brightness of the star. 

The application of the non-linear ephemeris yields a slightly larger scatter of the \bz\ measurements at close rotation phases compared to the constant period of, e.g., \citet{adelman:1997b}. Nevertheless, the overall distorted double-wave structure of the longitudinal field curve, \mybf{as well as satisfactory} agreement between the observations by \citet{thompson:1985} and \citet{donati:1997} obtained a decade apart, is retained.

\subsection{Circular polarization spectra}

An intensive spectropolarimetric observing campaign of \hd\ took place in 1994--1996 at the 6-m telescope of the Special Astrophysical Observatory (SAO). The star was observed in different wavelength bands with the Main Stellar Spectrograph (MSS) equipped with a CCD detector and \mybf{an achromatic circular polarization analyzer. Observations of \hd\ employed the second camera of the MSS, giving the effective spectral range of 5000--7000~\AA\ and a dispersion of 14~\AA/mm. From October 1994 to 1996 spectra were recorded using a $530\times580$ pixel detector, while a $1160\times1040$ CCD was used after March 1996. Both detectors provide a spectral resolution of $R\approx10\,000$, which is adequate for the analysis of rapid rotators such as \hd. ThAr lamp was used as a comparison source. The instrumental profile is close to a Gaussian with 2--2.5 pixel sampling of the FWHM.}

\mybf{
The extraction of the MSS Zeeman spectrograms was accomplished with the reduction package DECH20 by G. Galazutdinov\footnote{\url{http://www.gazinur.com/DECH-software.html}}. This software was also used to perform continuum normalization and to determine the wavelength solution from the measurement of the centroids of ThAr lines in the two orthogonally circularly polarized comparison spectra. Subsequent steps of the spectral processing and analysis employed the codes written in the ESO MIDAS environment \citep{kudryavtsev:2000}. Further details about spectropolarimetric observations of \hd\ as well as reduction and analysis of these data can be found in \citet{romanyuk:1998}.
}

\mybf{\hd\ was observed in three different wavelength regions during the spectropolarimetric campaign at SAO. By far} the most extensive data set of 30 circular polarization spectra is available for the region containing the \he\ line. These spectra 
\mybf{have a signal-to-noise ratio of 300--500 and cover the 5830--5900~\AA\ wavelength interval, which includes the sharp interstellar \ion{Na}{1}\,D lines. These features are not expected to show detectable polarization signal, allowing us to verify and fine-tune the relative wavelength alignment of the left- and right-hand circularly polarized beams in each observation.}

The rotation phases of the SAO spectra were calculated using the non-linear ephemeris of \citet{mikulasek:2008}, consistently with the longitudinal field data. These circular polarization observations provide a dense coverage of the rotational cycle of the star. The average phase sampling is 0.03 and the largest phase gap is 0.14 of the rotation period.

\citet{khokhlova:2000} used a subset of 9 spectrograms from the SAO \he\ data in their Doppler-Zeeman analysis of \hd. This subset provides a poor rotational phase coverage, with several phase gaps as large as 0.2--0.3.

\section{Multipolar modeling}
\label{multipolar}

In this section we aim to revise the multipolar model of \citet{bohlender:1994} using the new ephemeris of \hd\ and a more realistic value of the inclination angle of the stellar rotational axis. \citet{bohlender:1994} assumed $i=90\degr$, which together with the rotational period found in previous studies of \hd\ and \vsini\,=\,91~\kms\ determined below (Sect.~\ref{mdi}) implies an implausibly small radius, $R=2.8R_\sun$, for a main sequence B2 star. Other authors reported systematically larger radii: $R=3.7R_\sun$ \citep{shore:1990}, $R=4.3R_\sun$ \citep{zboril:1997}, $R=3.6R_\sun$ \citep{landstreet:2007}. This range of radii \mybf{corresponds to inclination angles} between 40\degr\ and 50\degr. Giving more weight to the more recent parameter determination by \citet{landstreet:2007}, we adopt $i=50\degr$. This inclination angle is close to $i=45\degr$ favored by the Doppler imaging analysis of \citet{khokhlova:2000}.

Keeping the inclination angle fixed, we used a least-squares optimization algorithm to fit the observed \bz\ curve with a low-order, axisymmetric multipolar magnetic field topology. The free parameters of this widely used global magnetic field parameterization \citep{landstreet:1988,landstreet:1989,landstreet:2000} include the angles $\beta$ and $\phi$, giving the obliquity and azimuth of the magnetic field axis. The field intensity and structure are determined by the polar strengths of the dipolar $B_{\rm d}$, quadrupolar $B_{\rm q}$, and octupolar $B_{\rm o}$ magnetic components.


\begin{figure}[!t]
\figps{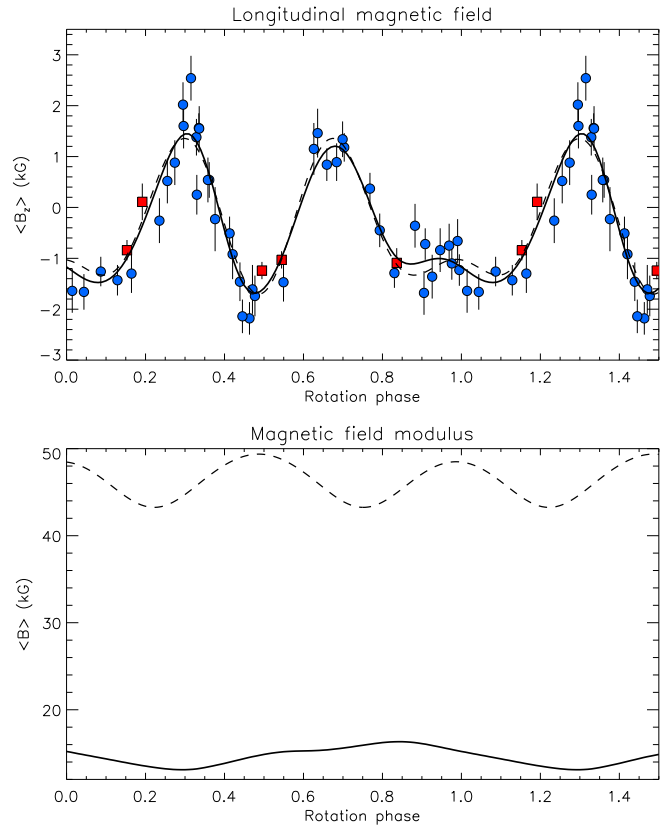}
\caption{\textit{Upper panel:} comparison of the mean longitudinal magnetic field of \hd\ measured by \citet{thompson:1985} (circles) and \citet{donati:1997} (squares) with the fit achieved by the axisymmetric multipolar model (dashed curve) and by the direct magnetic inversion (solid curve). \textit{Lower panel:} the mean field modulus variation predicted by the multipolar model (dashed curve) and by the magnetic DI field map (solid curve).}
\label{fig:bz}
\end{figure}

\begin{figure*}[!t]
\figps{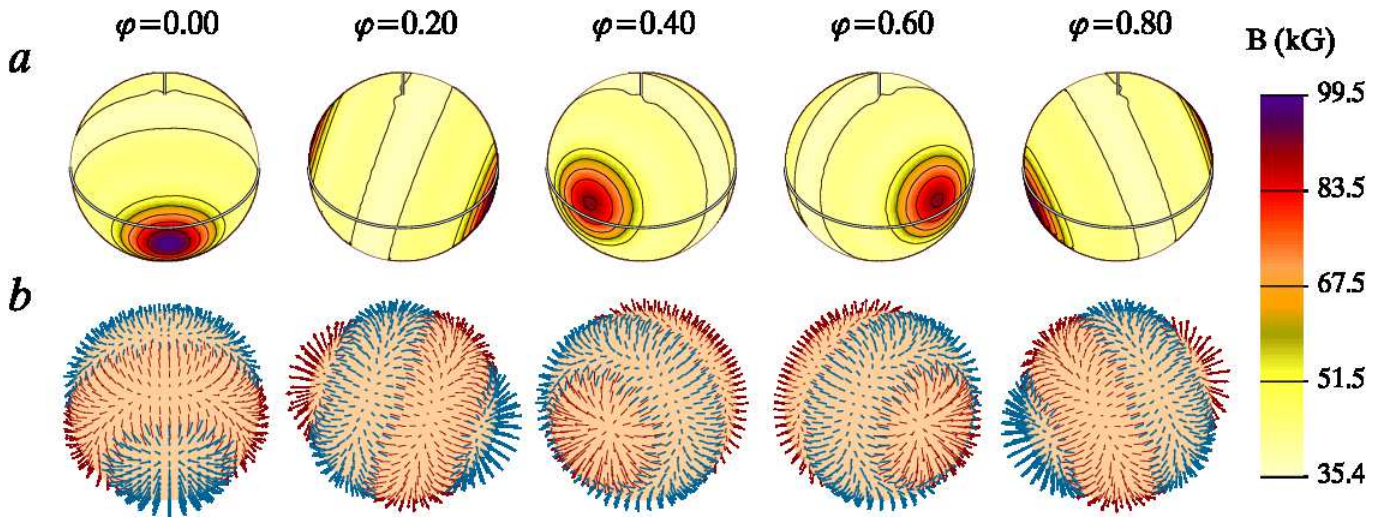}
\caption{Surface magnetic field structure \hd\ derived from multipolar fitting of the longitudinal field curve.
The star is shown at five equidistant rotational phases as indicated above the topmost row of spherical plots. The aspect corresponds to the inclination angle $i=50\degr$ and vertically oriented rotational axis.
{\bf a)} The distribution of the field strength, with contours of equal magnetic field strength plotted every 10~kG. The thick line shows the stellar rotational equator. Rotational axis is indicated by short the vertical bar.
{\bf b)} The orientation of the magnetic field vectors. In these vector maps the light arrows show the field vectors pointing outside the stellar surface and the dark arrows correspond to the vectors pointing inwards. The arrow length is proportional to the field strength.}
\label{fig:sph_bz}
\end{figure*}

Our best-fit multipolar model of the field topology in \hd\ is characterized by the reduce chi-square $\chi^2_\nu=1.26$ and has the following parameters: $i=50\degr$, $\beta=77.1\pm2.3\degr$, $\phi=184.8\pm1.7\degr$, $B_{\rm d}=-1.0\pm0.6$~kG, $B_{\rm q}=-4.3\pm13.6$~kG, $B_{\rm o}=97.0\pm7.3$~kG. The upper panel in Figure~\ref{fig:bz} shows the comparison between the \bz\ measurements and the model longitudinal field curve. Evidently a very strong axisymmetric octupolar magnetic field structure is capable of providing a reasonably good approximation of the complex longitudinal magnetic field variation of \hd\ (Figure~\ref{fig:bz}, upper panel). The inferred multipolar field topology is illustrated in more detail in Figure~\ref{fig:sph_bz}. These spherical maps demonstrate that the field strength varies between 35 and 100 kG over the stellar surface. The mean field modulus changes from 43 to 49~kG with the stellar rotation phase (Figure~\ref{fig:bz}, lower panel).

Taken at face value, the multipolar field modeling results suggest that \hd\ is indeed the most magnetized non-degenerate star.

\section{Magnetic Doppler imaging}
\label{mdi}

\begin{figure}[!th]
\fifps{6.85cm}{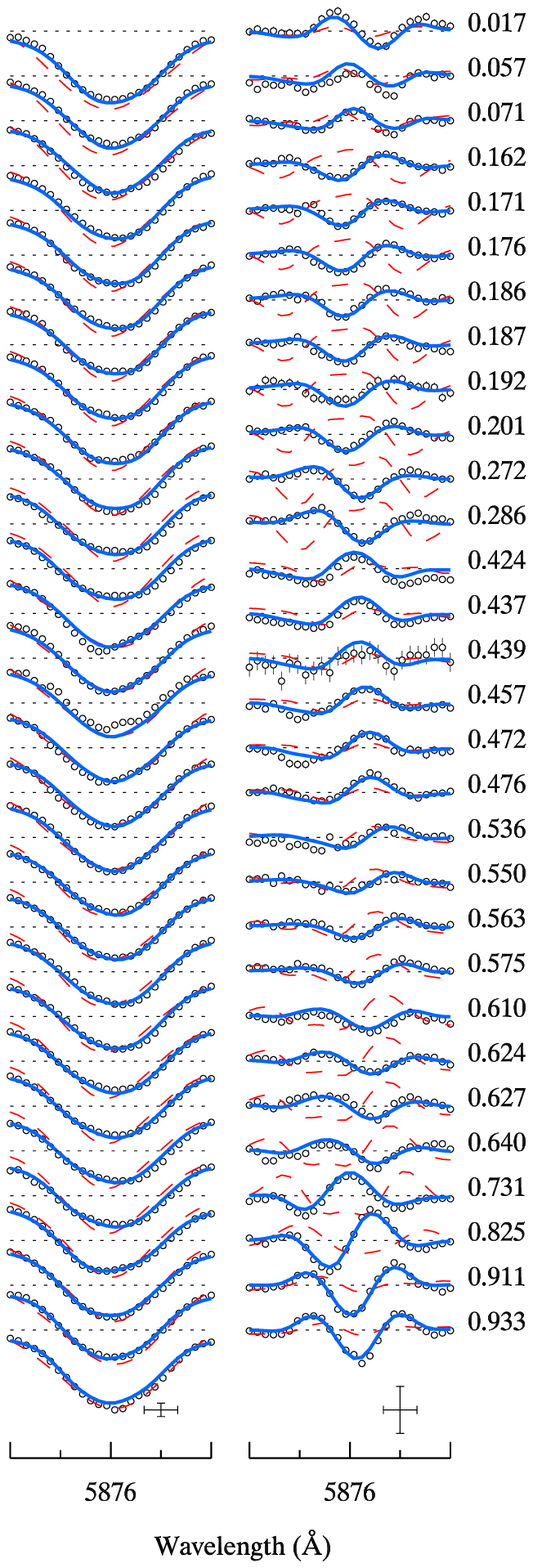}
\caption{Comparison of the observed Stokes $I$ (\textit{left}) and $V$ (\textit{right}) line profiles of \hd\ with the theoretical calculations corresponding to the best-fitting magnetic DI map (\textit{solid lines}) and to the axisymmetric multipolar field model shown in Figure~\ref{fig:sph_bz} (\textit{dashed lines}). The spectra for consecutive rotation phases are shifted vertically. The phases are indicated on the right. The bars at the bottom give the horizontal (1~\AA) and vertical (5\%) scales of the line profiles plots. \mybf{The horizontal dotted lines identify the continuum level for Stokes $I$ and the zero level for Stokes $V$.}}
\label{fig:prf}
\end{figure}

\begin{figure*}[!t]
\figps{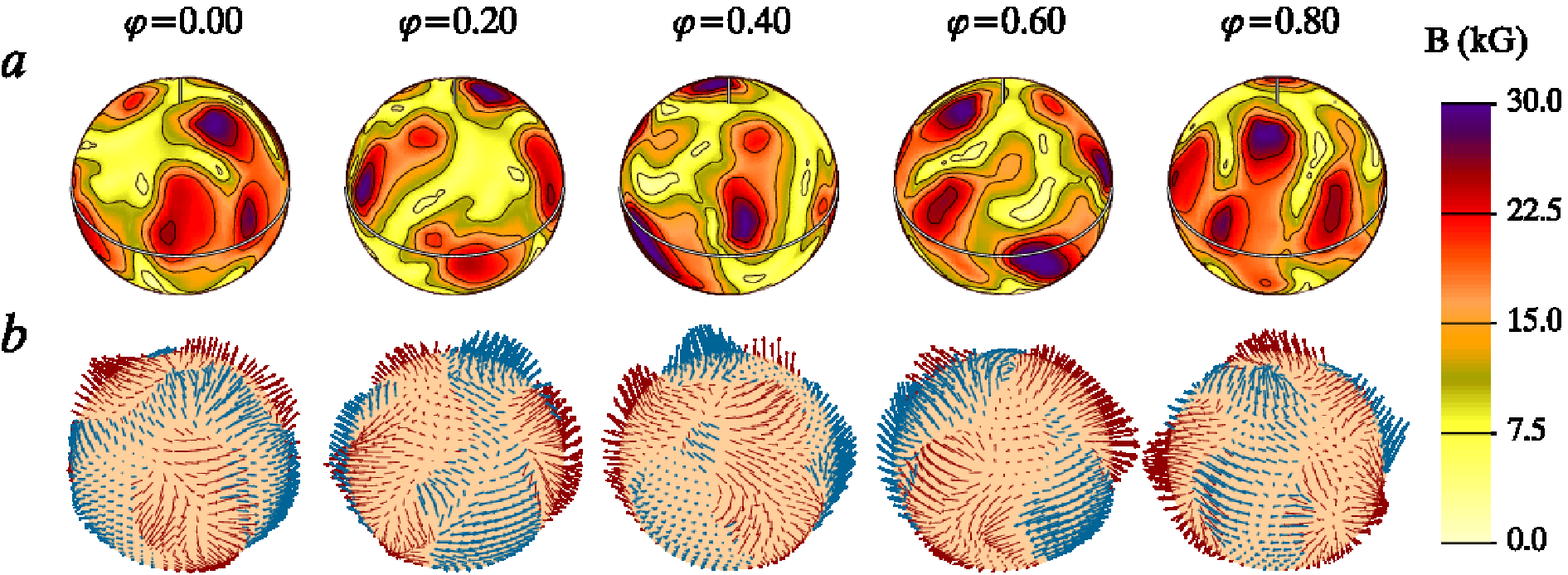}

\vspace*{0.5cm}

\figps{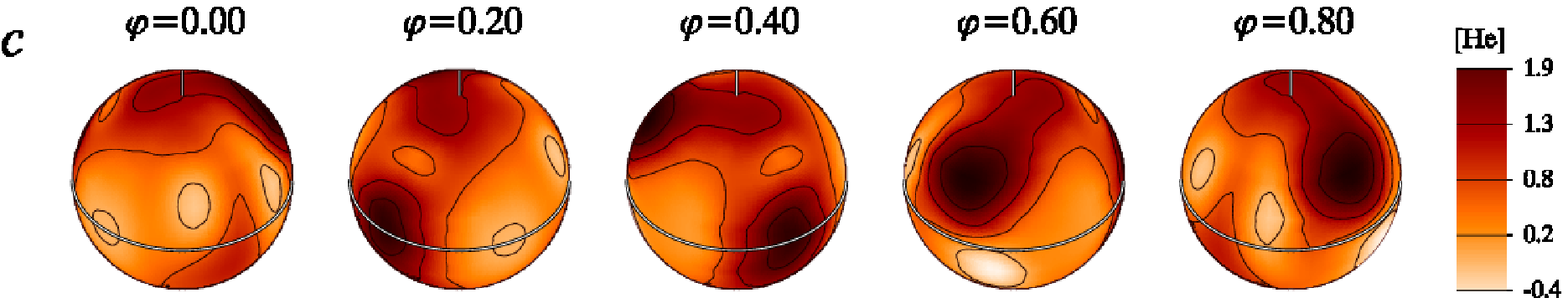}
\caption{Surface magnetic field structure and He distribution of \hd\ derived with magnetic Doppler imaging. 
The star is shown at five equidistant rotational phases as indicated above the first and third row of spherical plots. The aspect corresponds to the inclination angle $i=50\degr$ and vertically oriented rotational axis.
{\bf a)} The distribution of the field strength, with contours of equal magnetic field strength plotted every 5~kG. The thick line shows the stellar rotational equator. \mybf{The rotational axis} is indicated by the short vertical bar.
{\bf b)} The orientation of the magnetic field vectors. In these vector maps the light arrows show field vectors \mybf{pointing outward from the stellar surface} and the dark arrows correspond to the vectors pointing inwards. The arrow length is proportional to the field strength.
{\bf c)} The surface distribution of the He abundance, measured relative to the Sun, obtained simultaneously with the magnetic field geometry. The contour lines in the spherical abundance maps are plotted with a step of 0.5~dex.}
\label{fig:sph}
\end{figure*}

\begin{figure*}[!t]
\figps{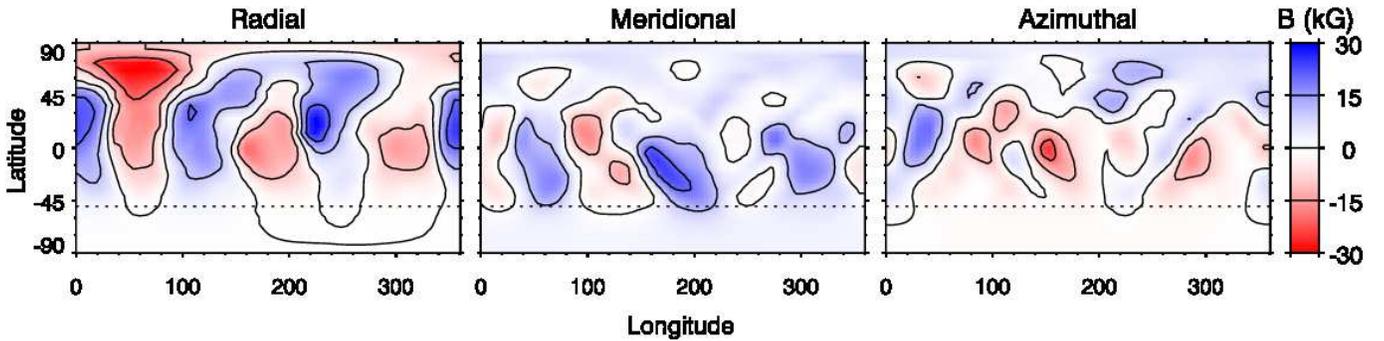}
\caption{Rectangular projection of the radial, meridional and azimuthal vector components of the magnetic field geometry of \hd\ derived in magnetic Doppler imaging.
The dotted line shows the lowest visible latitude for the adopted inclination angle $i=50\degr$. The lines of equal field strength are shown for the [$-30$,$+30$]~kG interval, with a 10~kG step.}
\label{fig:rec}
\end{figure*}

\begin{figure}[!t]
\fifps{6.5cm}{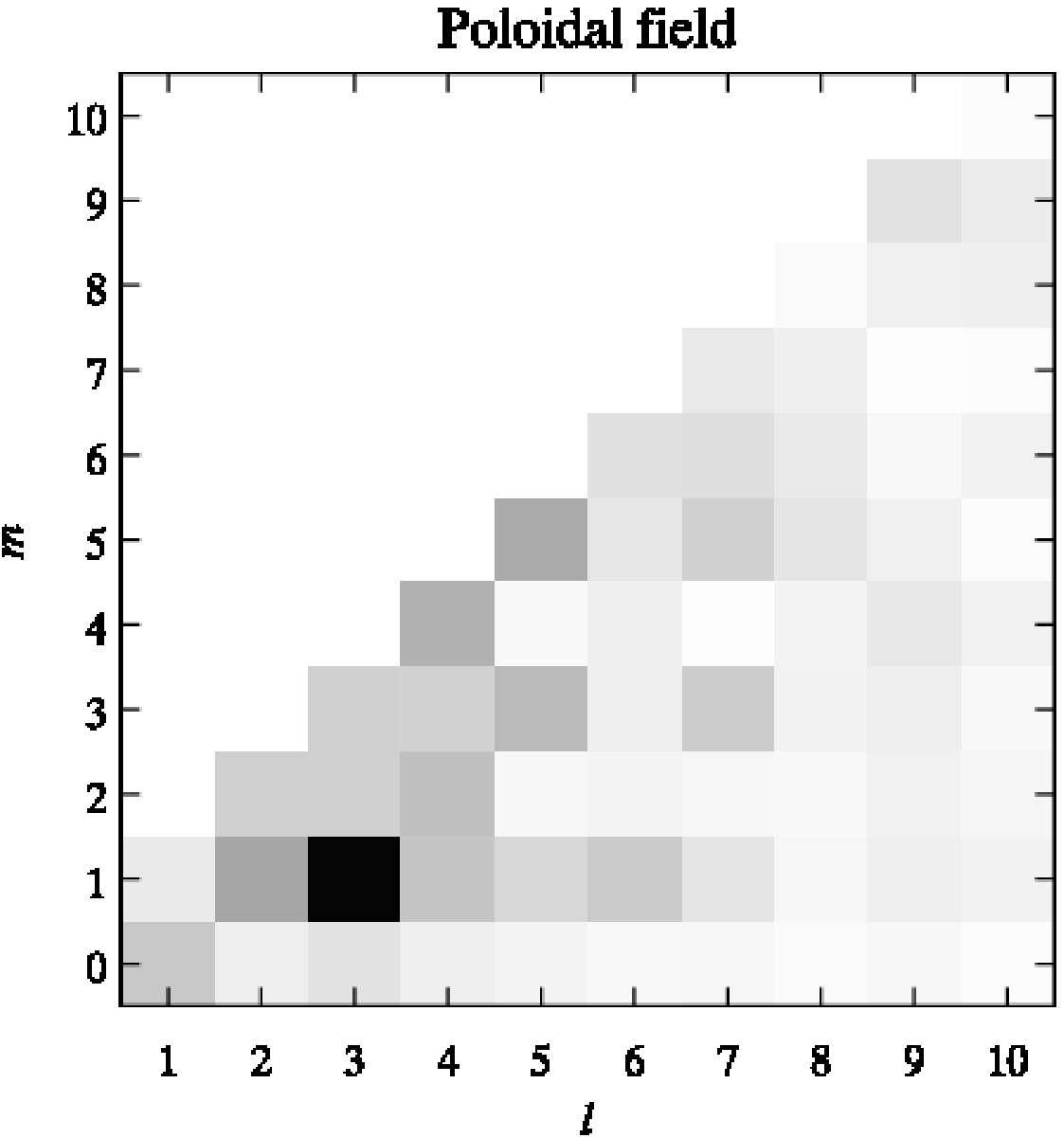}\vspace*{0.3cm}
\fifps{6.5cm}{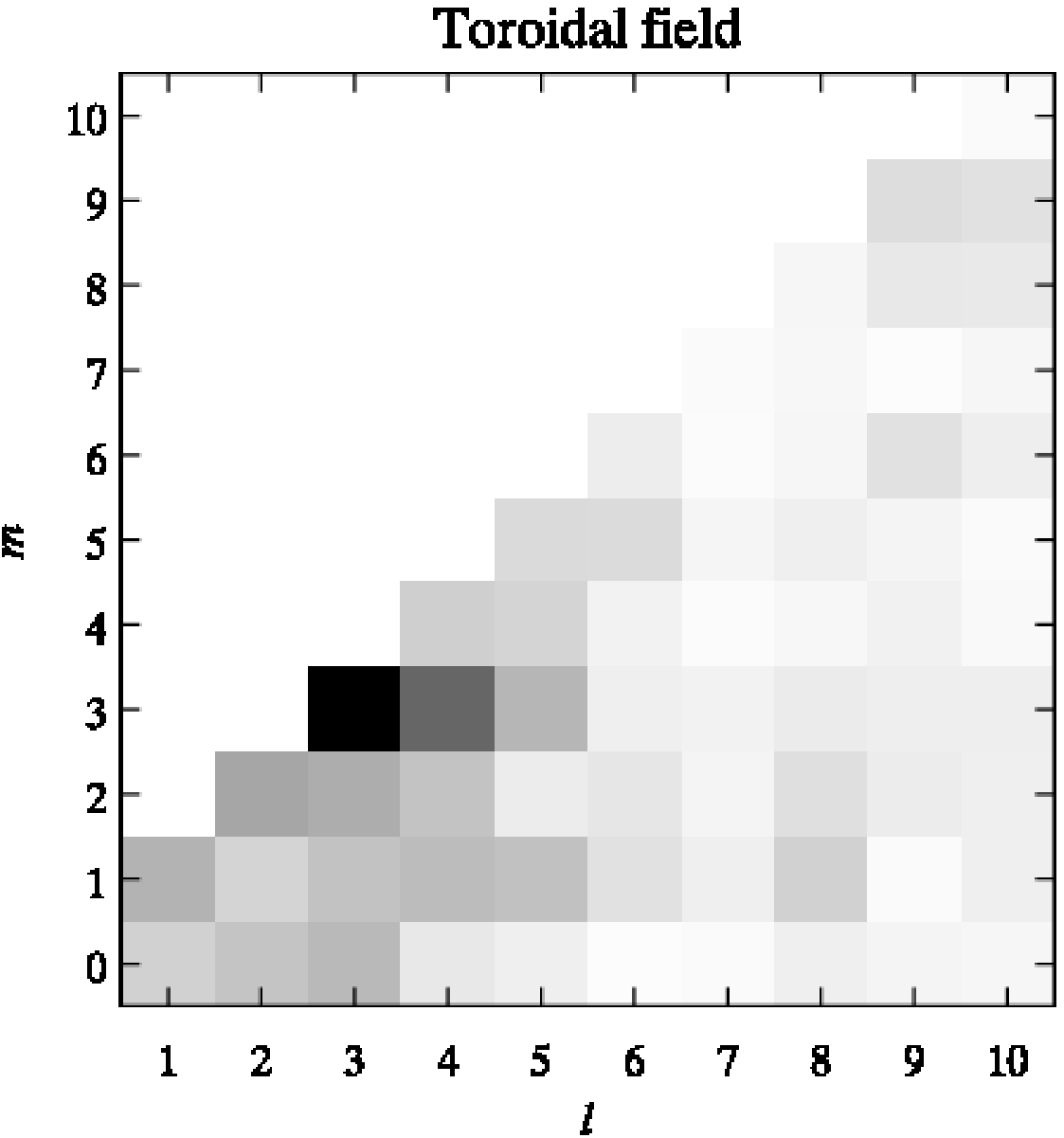}
\caption{Coefficients for the spherical harmonic expansion of the surface magnetic field structure of \hd. The top panel shows coefficients for the poloidal field, while the bottom panel corresponds to the toroidal field component.}
\label{fig:expansion}
\end{figure}

We reconstructed the magnetic field topology of \hd\ with the help of the Doppler imaging (DI) code \inv. This magnetic inversion code was described by \citet{piskunov:2002a} and \citet{kochukhov:2002c}, and applied to the analysis of Ap-star magnetic field geometries by \citet{kochukhov:2002b,kochukhov:2004d}, \citet{folsom:2008}, \citet{luftinger:2010}, and \citet{kochukhov:2010}. \inv\ determines the surface distributions of the magnetic field vector and chemical abundance from the phase-resolved spectropolarimetric observations. The code models individual line profiles using a realistic numerical treatment of the polarized radiative transfer in a stellar atmosphere. The surface maps are constrained by the Tikhonov regularization \citep{piskunov:2002a}, which ensures the uniqueness of the DI reconstruction and provides the smoothest solution justified by the observational data.

\mybf{The absence of the Stokes $Q$ and $U$ data in our study might complicate the magnetic field reconstruction, leading to some cross-talk between the radial and meridional field components at low latitudes \citep{kochukhov:2002c}. Furthermore, as suggested by the recent comparison of the magnetic inversions for the Ap star $\alpha^2$~CVn using four Stokes parameter observations and limiting the input data to circular polarization alone \citep{kochukhov:2010}, the linear polarization spectra enable reconstruction of small-scale surface magnetic features that cannot be detected with Stokes $V$ data. We emphasize that these problems arising due to lack of linear polarimetry are not unique to our study but apply to all previous Zeeman Doppler imaging analyses of cool active stars and hot massive stars \citep[e.g.,][]{donati:2003,donati:2006b}.}

For modeling \hd\ we used a version of our magnetic DI code that fits simultaneously the spectropolarimetric data and the longitudinal magnetic field curve \citep{kochukhov:2002b}. The surface maps were sampled on a 1176-element surface grid with an optimal distribution of surface zones \citep[see Figure~5 in][]{piskunov:2002a}. The polarized spectrum synthesis calculations for the \he\ line employed an LTE solar metallicity model atmosphere with \teff\,=\,22000~K and \logg\,=\,4.0 from the revised {\sc Atlas9} grid\footnote{\url{http://kurucz.harvard.edu/grids.html}}. The effective temperature and surface gravity adopted for \hd\ are consistent with the stellar parameters determined in previous studies of this star \citep{groote:1981,zboril:1997,hunger:1999,cidale:2007,landstreet:2007}.

Before proceeding to the magnetic inversion, we compared the predictions of the multipolar magnetic field model obtained in the previous section with the observed circular polarization profiles of the \he\ line. In this calculation the He abundance was adjusted to be $[He]\equiv\log(N_{\rm He}/N_{\rm H})-\log(N_{\rm He}/N_{\rm H})_\sun=-0.2$ to roughly match the average strength of this line in the unpolarized spectrum. The comparison of observations and multipolar model predictions is illustrated in Figure~\ref{fig:prf}. Evidently, the magnetic field topology inferred from the longitudinal field curve alone fails to account for the rotational phase dependence of the circular polarization within the \ion{He}{1} line. \mybf{In particular, major} disagreement between the predicted and observed Stokes $V$ profiles is found for the rotation phases 0.07--0.29 and 0.61--0.93. \mybf{Similar significant discrepancies between the details of observed $V$ profiles of magnetic CP stars and those predicted by multipolar models have been previously found for 53~Cam and $\beta$~CrB \citep{bagnulo:2001} but not for 78~Vir \citep{khalack:2006}.}

In contrast to the unsatisfactory results obtained with the multipolar field model, the DI reconstruction of the magnetic field geometry yields an excellent fit of the circular polarization spectra (Figure~\ref{fig:prf}) and an adequate description of the \bz\ curve (Figure~\ref{fig:bz}, upper panel). The resulting magnetic field distribution, constrained by the Tikhonov regularization but not limited to a low-order axisymmetric multipolar topology, is shown in Figures~\ref{fig:sph} and \ref{fig:rec}. The former figure illustrates the spherical surface maps of the field strength and orientation while the latter one presents the rectangular projections for the radial, meridional, and azimuthal components of the magnetic field vector. The He abundance distribution obtained together with the magnetic maps is shown in Figure~\ref{fig:sph}c. The He concentration changes from a moderate underabundance with respect to the solar chemical composition to an overabundance by up to a factor of 100 inside He spots.

Our magnetic DI reveals a strikingly different picture of the magnetic field in \hd\ compared to the suggestions of previous multipolar models. The field topology is complex and notably non-axisymmetric. There are six distinct areas of the positive and negative magnetic polarities, reminiscent of a somewhat distorted $\ell=3$ spherical harmonic. At the same time, the DI model yields a much weaker surface field strength than commonly assumed for \hd. The field intensity changes from $\approx$\,5~kG to $\approx$\,30~kG across the stellar surface, which is still a lot compared to a typical magnetic B-type star but certainly less than the field on several most extreme members of this group of stars. The mean magnetic field modulus of \hd\ determined from the DI maps varies between 13 and 16~kG. This is substantially lower than the $\approx$\,46~kG field modulus predicted by the multipolar model (see Figure~\ref{fig:bz}, lower panel).

Aiming to characterize more comprehensively the field topology obtained with the magnetic inversion, we applied to the final DI maps the spherical harmonic expansion method of \citet{piskunov:2002a}.  Both poloidal and toroidal spherical harmonic modes with the angular degree $\ell$\,=\,1--10 and all possible azimuthal numbers $m$ were included in the expansion. The resulting expansion coefficients are presented in Figure~\ref{fig:expansion}. We found that non-axisymmetric $\ell$\,=\,3--4 harmonic components give the largest contribution to the magnetic field topology of \hd, although a non-negligible power is evident for the lower-order multipoles and for the $\ell$ values up to 5--6. The relative fractions of the poloidal and toroidal field components appear to be similar.

\mybf{The presence of the significant toroidal field in \hd\ is an interesting and somewhat unexpected result. Although the toroidal component frequently dominates over the poloidal one in cool active stars \citep{donati:2003}, it is seldom diagnosed in hot magnetic stars and is never included in the parameterized low-order multipolar field models of these stars. Results of the present investigation as well as the previous polarization line profile studies by \citet{kochukhov:2004d} and \citet{donati:2006b} suggest that the toroidal component should not be disregarded in the magnetic field analysis of hot stars.}

\section{Magnetospheric structure}
\label{3dfld}

\begin{figure*}[!t]
\fifps{14cm}{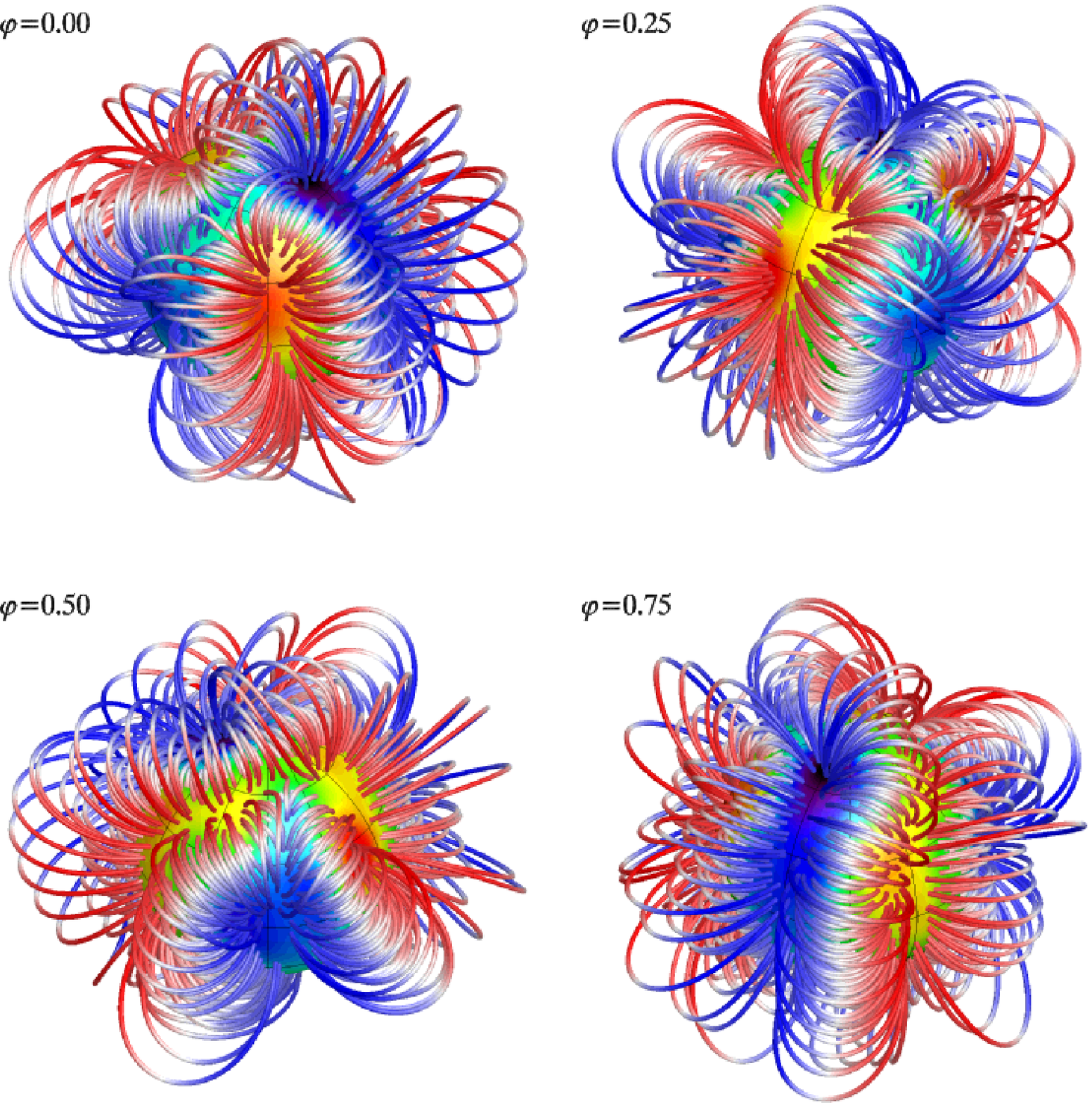}
\caption{Magnetospheric structure of \hd\ inferred with the potential field extrapolation of the surface field determined with magnetic DI. The star is shown at four rotation phases and the inclination angle $i=50\degr$. Only the field lines closed within a sphere of $R=2R_\star$ are shown here. The color along each field line and across the stellar surface changes according to the value of the radial magnetic field component.}
\label{fig:3d}
\end{figure*}

\mybf{The strong magnetic} fields of the early-type stars have a prominent effect on their radiatively driven winds and circumstellar environment. It is understood that a strong enough magnetic field confines the mass loss, producing the regions of hot, X-ray emitting \mybf{shocks and cooler}, denser clouds where the plasma is trapped by the magnetic field lines \citep{babel:1997,donati:2001,ud-doula:2002}. The magnetically confined wind and inhomogeneous magnetospheres manifest themselves with the characteristic variability of the wind-sensitive UV lines \citep{shore:1990}, emission in the hydrogen Balmer lines \citep{groote:1982}, and photometric indices \citep{townsend:2008}. The rotational spin-down, recently detected in some helium-strong stars including \hd\ \citep{mikulasek:2008,townsend:2010}, is also determined by the interaction between the wind and the stellar magnetic field \citep{ud-doula:2009}.

With the exception of the slowly rotating, chemically normal early B-type star $\tau$~Sco, which has a weak non-dipolar magnetic field \citep{donati:2006b}, the previous studies of the hot-star magnetospheres were limited to the objects with dipolar fields. Our analysis of \hd\ \mybf{probes a previously} unexplored parameter range, offering an insight into the magnetospheric structure and the wind-field interaction for a star possessing a strong magnetic field with a very complex geometry.

To investigate the extended magnetic field topology of \hd\ we calculated the three-dimensional structure of the circumstellar magnetic field using the potential field extrapolation technique of \citet{jardine:1999}. In this method the magnetospheric field is determined using the radial field component of the magnetic DI map and setting a spherical source function at the Alfv\'en radius, where the field lines are expected to become open under the wind pressure. This outer boundary surface is located where the wind confinement parameter $\eta$ \citep{ud-doula:2002}, characterizing the ratio between the magnetic field energy and the kinetic energy of the wind, is on the order of unity.

Adopting the typical magnetic field strength at the stellar surface $B_\star$\,$\approx$\,14.5~kG, the stellar radius $R_\star$\,$\approx$\,4$R_\sun$ \citep{shore:1990,zboril:1997,mikulasek:2008}, the average mass-loss rate $M$\,$\sim$\,$10^{-9}M_\sun$\,year$^{-1}$, and the terminal wind speed $v_\infty$\,$\sim$\,$10^{3}$~\kms\ \citep{krticka:2006,mikulasek:2008}, we find the wind confinement parameter $\eta$\,$\sim$\,$3\times10^6$ for \hd. This yields the Alfv\'en radius $R_{\rm A}>40R_\star$ for a dipole-dominated surface field and $R_{\rm A}>10R_\star$ for a quadrupolar magnetic topology.

Despite several uncertainties in our estimate of $\eta$ and $R_{\rm A}$, the results of the potential field extrapolation in the immediate vicinity of the star are weakly dependent on the assumed location of the source surface as long as $R_{\rm A} \gg R_\star$. Therefore, we limit the following discussion to the inner $2R_\star$ of the stellar magnetosphere. The circumstellar magnetic field topology established  in this region with the potential field extrapolation procedure is illustrated in Figure~\ref{fig:3d}. In this plot the closed magnetic field lines are rendered for four rotation phases. The color variation along each field line corresponds to the radial field component. 

Not unexpectedly, we find that the magnetospheric field structure of \hd\ is far more complex than a dipolar or an axisymmetric low-order multipolar field. The extended field topology is dominated by a network of the closed magnetic field lines, concentrated along the six ridges stretched mostly in the meridional direction. The location of the closed field line loops is determined by the six distinct areas of the positive and negative field polarity on the stellar surface (see Figure~\ref{fig:rec}). This non-trivial inner magnetospheric field structure should give rise to an equally complex rotational phase dependence of the parameters tracing the wind inhomogeneities and circumstellar matter.

\section{Discussion}
\label{disc}

\begin{figure}[!t]
\fifps{8cm}{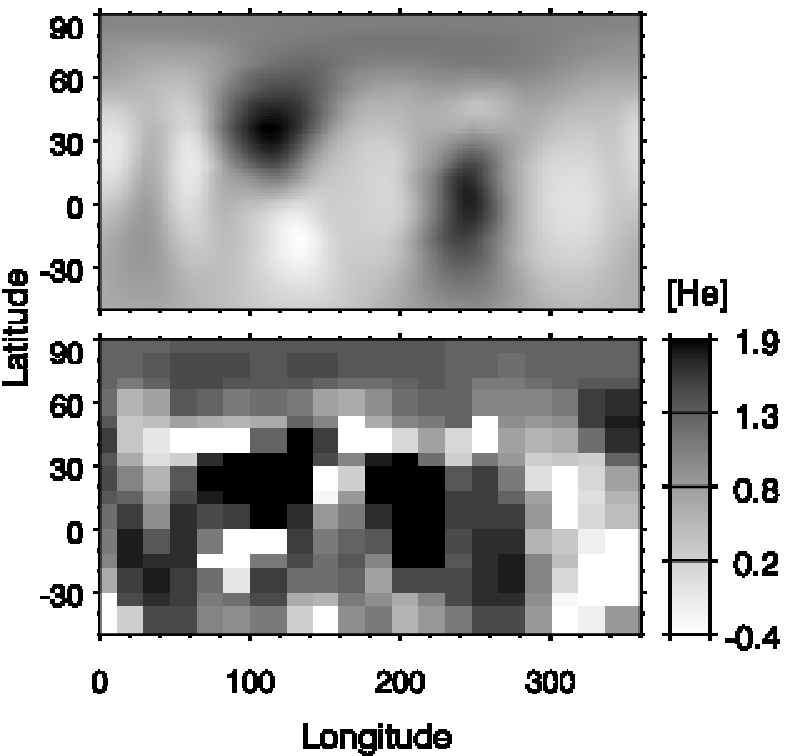}
\caption{Comparison of the He abundance distribution derived for \hd\ in our study (\textit{top panel}) and by \citet{khokhlova:2000} (\textit{bottom panel}).}
\label{fig:he}
\end{figure}

In this paper we presented a new analysis of the magnetic field observations of the unique He-strong star \hd. The unusually complex magnetic field of this object distinguishes it from other early-type stars, which have mostly dipolar field topologies. We reanalyzed the longitudinal magnetic field measurements of \hd, finding that a very strong, $\sim100$~kG, octupolar field is required to reproduce the rotational modulation of \bz\ under the assumption of an axisymmetric multipolar topology. However, we found that this multipolar model is incompatible with the circular polarization time-series observations of the \he\ line. Using both the longitudinal magnetic field measurements and the Stokes $V$ spectra, we derived a new model of the surface magnetic field topology of \hd\ with the help of magnetic Doppler imaging. This magnetic field structure for the first time explained \mybf{all the different types} of magnetic observations available for \hd. The field turns out to be highly complex, bearing no resemblance to a low-order, axisymmetric multipolar geometry commonly assumed for early-type stars. Thus, together with $\tau$~Sco, which has a much weaker field \citep{donati:2006b}, \hd\ belongs to an unusual group of the early-type stars with conspicuously non-dipolar fields.

Our investigation of the longitudinal magnetic field variations of \hd\ clearly demonstrated the fundamental limitation of the multipolar field models based on fitting the \bz\ curve alone. While the multipolar model requires a record-strong field to reproduce the \bz\ measurements, magnetic DI yields an order of magnitude weaker magnetic field intensity, with the average field strength $\approx14.5$~kG. This is below the field modulus directly measured with the Zeeman splitting in several Ap/Bp stars \mybf{\citep{babcock:1960,babel:1995,mathys:1997b,hubrig:2005,kochukhov:2006c,ryabchikova:2006a,freyhammer:2008}}.

\mybf{The He abundance distribution in \hd\ was previously studied by \citet{khokhlova:2000} using a subset of data analyzed here and a somewhat less sophisticated line profile modeling technique \citep{vasilchenko:1996}. Their He abundance map is compared with ours in Figure~\ref{fig:he}. The He spot distribution derived by Khokhlova et al. has a significantly higher contrast and abundance range, but the location of the two main overabundance features agree reasonably well with our results. On the other hand, the fine structure of the He variation across the surface of \hd\ inferred by Khokhlova et al. is likely to be spurious.}

Using the magnetic field topology inferred with Doppler imaging as a starting point, we applied the potential field extrapolation method to calculate the extended magnetospheric field geometry of \hd. The field structure above the stellar surface was found to be complex and non-axisymmetric, with most of the field lines closed within a sphere of $2R_\star$. This magnetic geometry must correspond to an unusually complicated configuration of the magnetospheric plasma, which is broadly consistent with \citet{shore:1990} observations of the complex variation of the \ion{C}{4} and \ion{Si}{4} UV wind diagnostic lines.

The existence and stability (at least on the time-scale of a few decades covered by the longitudinal field measurements) of the very complex magnetic field in \hd\ poses an interesting challenge to the fossil field theory, which is usually applied to explain the magnetism of massive stars. Numerical simulations by \citet{braithwaite:2006} demonstrated that both toroidal and poloidal magnetic fields of comparable strength must be present in the stellar interior to maintain the global stability of the fossil field structure. On the surface, the field appears to be purely poloidal, axisymmetric and dipolar-like. In the context of these results, it is difficult to explain complex non-dipolar magnetic fields with a substantial toroidal contribution at the stellar surface. 

Subsequent MHD calculations by \citet{braithwaite:2008} revealed a different class of the stable magnetic equilibrium solutions, consisting of one or several large-scale twisted flux tubes lying under the surface of the star. Each tube is formed by the toroidal field, with the poloidal field lines wrapped around it. The multiple flux tube scenario, realized in Braithwaite's simulations if the initial unstructured magnetic flux is sufficiently spread out inside the star, seems to be applicable to the cases of $\tau$~Sco and \hd. However, the theory still has difficulties explaining the observed presence of significant surface toroidal field in both stars. \mybf{In particular, it is challenging to understand the origin and nature of the current system in the stellar atmosphere required to support these toroidal fields.}

The complex equilibrium fields studied by \citet{braithwaite:2008} evolve on diffusive time-scales, comparable to the stellar life-time on the main sequence. The flux tubes gradually rise towards the surface and undergo lengthening and narrowing due to the loss of toroidal magnetic flux. It is generally expected that the diffusive field decay proceeds faster for higher-order magnetic components. In this case, the magnetic topology of the old stars should be closer to dipolar, while the field structure of young stars, such as \hd\ \citep[$\log t=6.55$ yr,][]{landstreet:2007}, could be more complex. However, \citet{braithwaite:2008} presented examples of simpler axisymmetric fields evolving into more complex, non-axisymmetric ones. A calculation tuned for the specific parameters of \hd\ is required to ascertain whether these changes can be observable and to what extent the unusually high degree of complexity of the field in this star can be related to its youth. Nevertheless, we can conclude that the presence of a very complex strong magnetic field in a young, early B-type star is not inconsistent with the current qualitative predictions of the fossil field theory.

\acknowledgements
We thank Dr. Z. Mikul\'a\v sek for his help with the rotational phase calculation for \hd.
O.K. is a Royal Swedish Academy of Sciences Research Fellow supported by grants from the Knut and Alice Wallenberg Foundation and the Swedish Research Council.


\end{document}